\documentclass[12pt]{article}
\usepackage{graphicx}

\def\gtrsim{\stackrel{>}{{}_\sim}}
\def\lesssim{\stackrel{<}{{}_\sim}}

\newcommand\pubnumber{Article 45 in eConf C1304143}
\newcommand\pubdate{\today}

\def\Title#1{\begin{center} {\Large #1 } \end{center}}
\def\Author#1{\begin{center}{ \sc #1} \end{center}}
\def\Address#1{\begin{center}{ \it #1} \end{center}}

\newcommand\pubblock{\rightline{\begin{tabular}{l} \pubnumber\\
         \pubdate  \end{tabular}}}
\newenvironment{Abstract}{\begin{quotation}  }{\end{quotation}}
\newenvironment{Presented}{\begin{quotation} \begin{center} 
             PRESENTED AT\end{center}\bigskip 
      \begin{center}\begin{large}}{\end{large}\end{center} \end{quotation}}
\def\Acknowledgements{\bigskip  \bigskip \begin{center} \begin{large}
             \bf ACKNOWLEDGEMENTS \end{large}\end{center}}

\def\beq{\begin{equation}}
\def\eeq#1{\label{#1}\end{equation}}
\def\eeqn{\end{equation}}

\def\beqa{\begin{eqnarray}}
\def\eeqa#1{\label{#1}\end{eqnarray}}
\def\eeqan{\end{eqnarray}}

\let\bar=\overbar

\def\Dslash{\not{\hbox{\kern-4pt $D$}}}
\def\dslash{\not{\hbox{\kern-2pt $\del$}}}

\def\msb{{\bar{\ssstyle M \kern -1pt S}}}

\begin{document}
\begin{titlepage}
\pubblock

\vfill
\Title{Polarization of GRB Prompt Emission}
\vfill
\Author{Kenji Toma}
\Address{Department of Earth and Space Science, Osaka University, Toyonaka 560-0043, Japan\\
toma@vega.ess.sci.osaka-u.ac.jp}
\vfill
\begin{Abstract}
We review the recent observational results of the gamma-ray linear polarization of 
Gamma-Ray Bursts (GRBs), and discuss some theoretical implications for the prompt emission
mechanism and the magnetic composition of GRB jets. We also report a strict observational 
verification of $CPT$ invariance in the photon sector as a result of the GRB polarization
measurements.
\end{Abstract}
\vfill
\begin{Presented}
GRB 2013 \\
the 7th Huntsville Gamma-Ray Burst Symposium\\
Nashville, Tennessee, USA, April 14--18, 2013
\end{Presented}
\vfill
\end{titlepage}
\def\thefootnote{\fnsymbol{footnote}}
\setcounter{footnote}{0}

\section{Introduction}

Gamma-Ray Bursts (GRBs) are brief, intense flashes of gamma-rays originating at cosmological distances,
and they are the most luminous objects in the universe. The emitted radiation energy is dominant in the
0.1--1 MeV energy range.
It has been established that the prompt gamma-ray
emission is produced in the relativistic jets. However, in spite of extensive observational and theoretical
efforts, several key questions concerning the nature of the central engines of the relativistic jets and 
the jets themselves remain poorly understood (for recent reviews, see \cite{gehrels12,sinoue13,amati13}).
In fact, some of these questions are very difficult or
even impossible to answer with the light-curve and spectral information currently collected from the 
optical to GeV gamma-rays. 
On the other hand, polarization information can lead to unambiguous answers to these questions. 
In particular, polarimetric observations of GRBs can address the prompt emission mechanism, magnetic 
composition, and geometric structure of GRB jets \cite{toma09,lazzati06}.
The polarimetric data taken together with the light-curve and spectral data may reveal the driving mechanism
of the jets and the nature of the central engines.

Recently, we have measured gamma-ray polarizations in the 70--300 keV range 
with $\sim 3\sigma$ confidence levels, for the first time, in three bursts 
with the Gamma-ray Burst Polarimeter (GAP) onboard the small solar-power-sail demonstrator IKAROS 
\cite{yonetoku11,yonetoku12}. We review the observational results of the GAP as well as other instruments
with discussing some theoretical implications. 
We also report a strict observational verification of $CPT$ invariance in the photon sector as a result of
the GRB polarization measurements \cite{toma12}.

\section{Observations}

There were several reports of detections of linear polarization with low significance. Coburn \& Boggs (2003) 
\cite{coburn03} reported detection of strong polarization from GRB 021206 with RHESSI solar satellite.
However, independent authors analyzed the same data, and concluded that no polarization signals were 
confirmed \cite{rutledge04,wigger04}. INTEGRAL-SPI and -IBIS data showed detections of polarization with
$\sim 2\sigma$ confidence level from GRB 041219 \cite{kalemci07,mcglynn07,gotz09}. However, the results
of SPI and IBIS for the brightest pulse of GRB 041219 appear inconsistent with each other, i.e.,
the SPI teams detected strong polarization of $\Pi = 98 \pm 33\%$ and $\Pi = 63^{+31}_{-30}\%$ with $2\sigma$
statistical level, while the IBIS team reported a strict upper limit of $\Pi < 4 \%$ (although their results 
for the other temporal intervals are consistent). Therefore, the previous
reports of the gamma-ray polarimetry for GRBs are all controversial.

IKAROS is a small solar-power-sail 
demonstrator \cite{kawaguchi08}, and successfully launched on 21 May 2010. The GAP onboard IKAROS
is fully designed to measure linear polarization in prompt emission of GRBs in the energy range of 
70--300 keV. The GAP's high axial symmetry in shape and high gain 
uniformity are keys for reliable measurement of polarization and avoiding fake modulation due to 
background gamma-rays. These realized the quite small systematic uncertainty of $\simeq 1.8\%$ level
\cite{yonetoku11PASJ}.

\begin{table}[t]
\begin{center}
\begin{tabular}{|l|cccc|}  
\hline
Event name   &    $\Pi$    & $2\sigma$ limit & Detection significance &  PA change  \\
\hline 
GRB 100826A & $27 \pm 11$\% & $>6\%$ & $2.9\sigma$ & yes \\
GRB 110301A & $70 \pm 22$\% & $>31\%$ & $3.7\sigma$ & no \\
GRB 110721A & $84^{+16}_{-28}$\% & $>35\%$ & $3.3\sigma$ & no  \\ 
\hline
\end{tabular}
\caption{Polarimetric data of the three GRBs obtained with GAP. The Polarization degrees
$\Pi$ are shown with $1\sigma$ error. The `2$\sigma$ limit' means the lower limit on $\Pi$
at the $2\sigma$ statistical significance level. The `detection significance' means the significance
levels for $\Pi > 0\%$.}
\label{tab:data1}
\end{center}
\end{table}

\begin{table}[t]
\begin{center}
\begin{tabular}{|l|ccc|}  
\hline
Event name  &  $T_{90}$ [s] & fluence [${\rm erg}\; {\rm cm}^{-2}$] &  $E_p$ [keV]  \\
\hline 
GRB 100826A & $\simeq 150$ & $(3.0\pm 0.3)\times 10^{-4}$ & $606^{+134}_{-109}$ \\
GRB 110301A & $\simeq 5 $ & $(3.65\pm 0.03)\times 10^{-5}$ & $106.8^{+1.85}_{-1.75}$    \\
GRB 110721A & $\simeq 24$ & $(3.52\pm 0.03)\times 10^{-5}$ & $393^{+199}_{-104}$   \\ 
\hline
\end{tabular}
\caption{Light-curve and spectral data of the three GRBs taken from the GCN circulars.
$E_p$ is the photon energy of the time-averaged $\nu F_{\nu}$ spectrum.}
\label{tab:data2}
\end{center}
\end{table}

The GAP detected the linear polarization of the prompt emission of GRB 
100826A, GRB 110301A, and GRB 110721A. The polarimetric data as well as the light-curve and
spectral data of these three bursts are summarized in Table~\ref{tab:data1} and \ref{tab:data2}.
The polarization degrees $\Pi > 0\%$ at $\sim 3 \sigma$ confidence level, and these are
the most convincing detections of polarization of GRB prompt emission so far.
See Yonetoku et al. (2011; 2012) and Toma et al. (2012) \cite{yonetoku11,yonetoku12,toma12} 
for more details on the data analysis.

We see that there are cases with and without a significant change of the polarization angle (PA).
GRB 100826A, with long duration $T_{90} \sim 100\;$s, shows a PA change, while GRB 110301A
and 110721A, with short duration $T_{90} \sim 10\;$s, shows no PA change.
On the other hand, the polarization is detected both for the GAP observed energy range
$< E_p$ (GRB 100826A and GRB 110721A) and for $>E_p$ (GRB 110301A). The time-averaged fluxes
(the fluences divided by $T_{90}$) of the three bursts are all 
$\sim 3\times 10^{-6}\;{\rm erg}\; {\rm cm}^{-2}\; {\rm s}^{-1}$, which is very high.
We note that no spectroscopic redshifts were determined for these three bursts.

The polarimetric data of GRB 041219A as well as the recent report on GRB 061122 
with IBIS onboard INTEGRAL appear consistent with the GAP results listed above. GRB 041219A shows PA 
changes, and GRB 061122 has $\Pi \gtrsim 30\%$ at $2\sigma$ significance level 
\cite{gotz09,gotz13}.

\section{Implications for Emission Mechanism}

Here we present some theoretical implications for the emission mechanism, focusing on the 
above results that there are cases with and without a significant PA change, and that
$\Pi > 30\%$ for the cases of no PA change. 

For the light-curve and spectral dataset of GRB prompt emission in the 0.1--1 MeV energy range, 
the synchrotron emission model and the photospheric quasi-thermal emission model
have been actively debated (for reviews, see \cite{gehrels12,sinoue13,toma11}).
In general, the relativistic jets are optically thick in the vicinity of the central engine,
and the photospheric emission is released at the photosphere. The jets are optically thin
outside the photosphere and can emit the synchrotoron radiation through the internal shock
dissipation and/or the magnetic energy dissipation. The long-standing problem is which emission
is dominant in the 0.1--1 MeV energy range, synchrotron or photospheric. 

\subsection{Synchrotron Models}

First, let us discuss the synchrotron models. They include three different models in respect of 
the magnetic field structure in the emitting region. For all the models, we assume that the electrons 
have an isotropic pitch angle distribution and a non-thermal energy spectrum, and thus the polarization
degree from a local emitting point is $\Pi^{\rm syn}_{\rm max} = -\gamma_{\rm B}/(-\gamma_{\rm B}+2/3)$, where
$\gamma_{\rm B}$ is the photon power-law index \cite{rybicki79}.  
Typical bursts have $\gamma_{\rm B} \simeq -1$ for $E < E_p$ and
$\gamma_{\rm B} \lesssim -2$ for $E > E_p$.

\subsubsection{SO model}

One type of the magnetic field structure is a helical field, 
which may be advected from the central engine. Such globally ordered fields 
can produce high-$\Pi$ emission \cite{lyutikov03,granot03,toma09}. We call this the ``SO model''
(synchrotron, ordered-field model).
As the simplest model, we may assume an instantaneous emission from a thin spherical shell moving 
radially outward with a bulk Lorentz factor $\Gamma \gg 1$ and an opening angle $\theta_j$, where
the emissivity is uniform over the shell.

\begin{figure}[t]
\begin{minipage}{0.5\hsize}
\begin{center}
\includegraphics[scale=0.6]{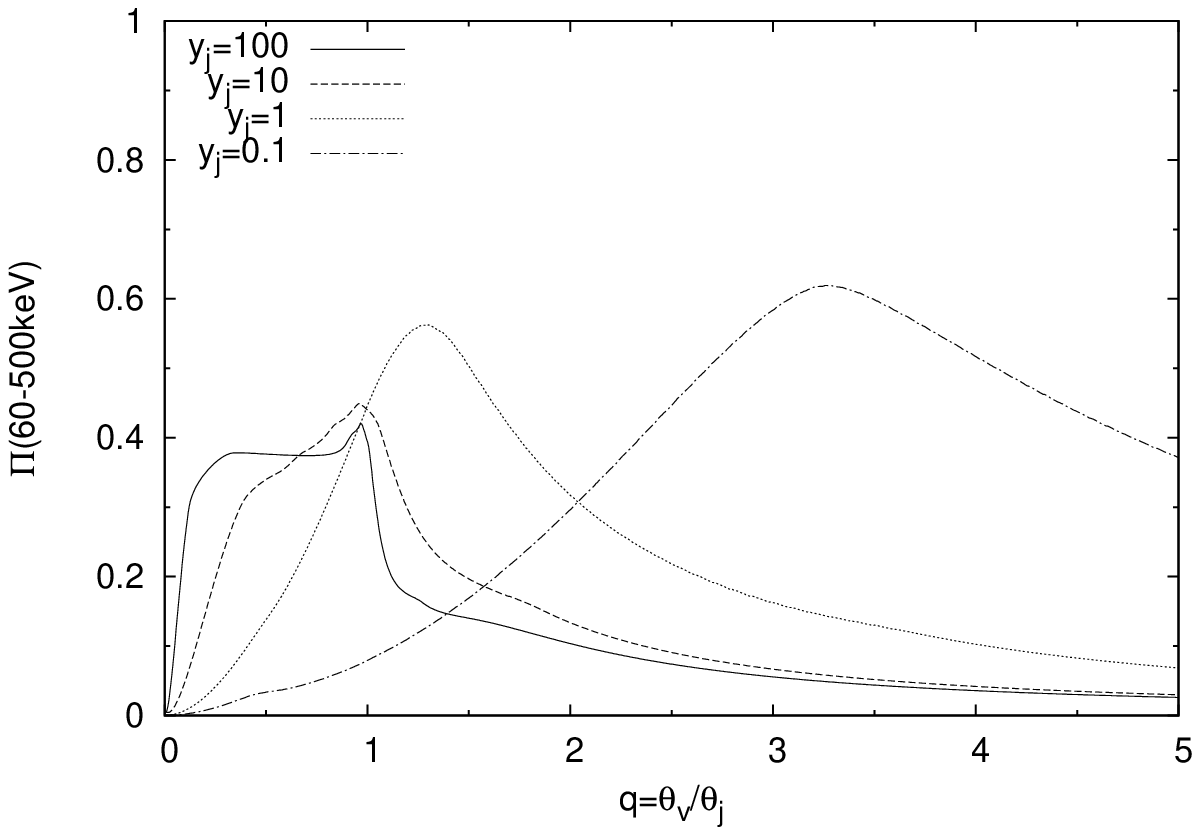}
\end{center}
\end{minipage}
\begin{minipage}{0.5\hsize}
\begin{center}
\includegraphics[scale=0.6]{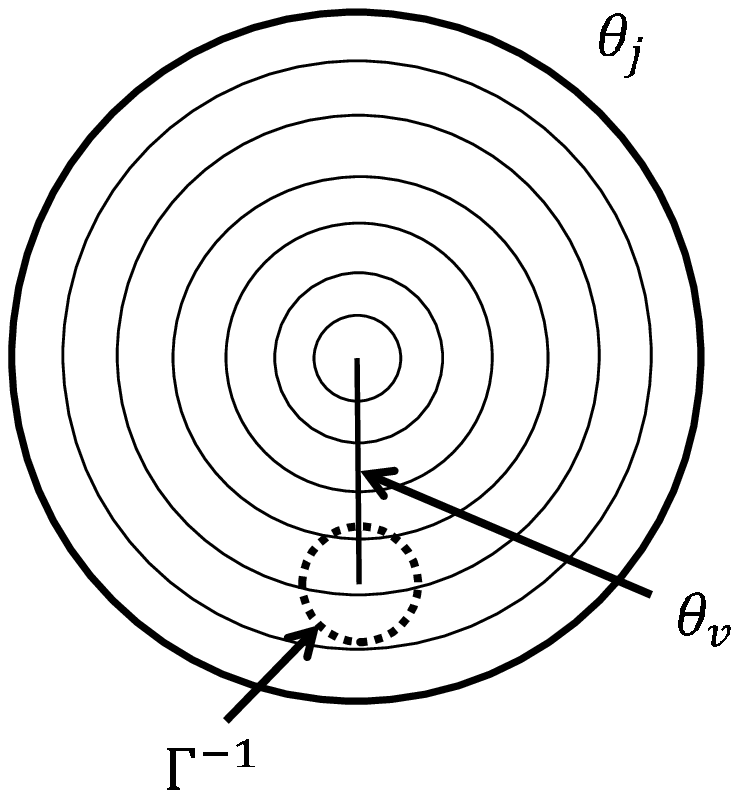}
\end{center}
\end{minipage}
\caption{
Left: Polarization degrees as functions of $q=\theta_v/\theta_j$ in the SO model, where 
$\theta_v$ is the viewing angle of the line of sight and $\theta_j$ is the jet opening angle.
$y_j \equiv (\Gamma \theta_j)^2$, where $\Gamma$ is the bulk Lorentz factor of the jet.
Typical parameters are adopted for the emission spectrum (see \cite{toma09} for details). 
Right: Schematic picture of the jet with the toroidal component of the magnetic fields (thin lines). 
Only a fraction of the emitting shell, 
$\theta < \Gamma^{-1}$ around the line of sight is bright because of the relativistic beaming effect.
}
\label{fig:SO}
\end{figure}

Figure~\ref{fig:SO} (left) shows $\Pi$ calculated in this model as a function of the viewing angle 
$\theta_v$ in respect of the jet axis for different values of $y_j \equiv (\Gamma \theta_j)^2$.
Since we consider the spherical shell, only a fraction of the shell with $\theta < \Gamma^{-1}$
around the line of sight is bright because of the relativistic beaming effect. This bright region
is small for the cases of $\Gamma^{-1} \ll \theta_j$ (i.e., $y_j \gg 1$) (see Fig~\ref{fig:SO} right). 
For the on-axis case, i.e., $\theta_v < \theta_j$, the direction of the 
magnetic field is quite ordered in the bright region, one has high polarization degree, 
$\Pi \sim 40\%$. 

\begin{figure}[t]
\begin{center}
\includegraphics[scale=0.6]{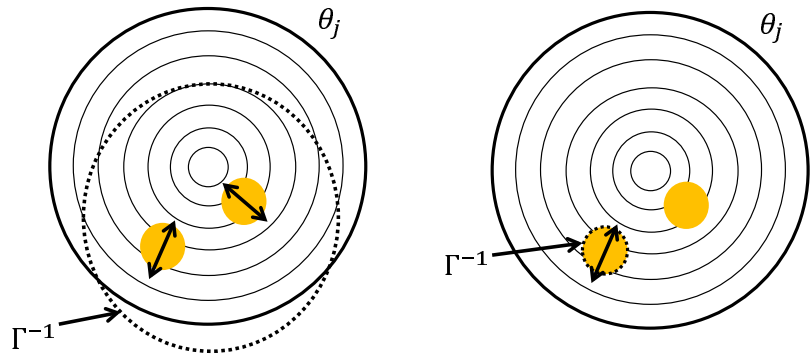}
\end{center}
\caption{
Schematic picture of the patchy emission in the SO model. The left and right ones correspond
to the cases of $\Gamma^{-1} \sim \theta_j$ and $\Gamma^{-1} \ll \theta_j$, respectively.
The thick arrows represent the polarization vectors.
}
\label{fig:SOpatch}
\end{figure}

Another important point is that the PA does not vary for a fixed $\theta_v$ but different $\Gamma$.
Therefore, the observed significant PA change may suggest that the emission is not uniform over the
shell, but consists of multiple patches with characteristic angular size much smaller than jet
opening angle, $\theta_p \ll \theta_j$ \cite{yonetoku11} (see Figure~\ref{fig:SOpatch}). 
In the case of $\Gamma^{-1} \sim \theta_j$,
it is natural that one sees multiple patches with different magnetic field directions, and 
observes significant PA changes. On the other hand, if $\Gamma^{-1} \ll \theta_j$, one only sees
a limited range of the curved magnetic fields, which leads to no significant PA change even if the 
emission is patchy. In such a scenario, GRB 100826A corresponds to the case of $\Gamma^{-1} \sim \theta_j$,
while the other two bursts with no PA change correspond to the case of $\Gamma^{-1} \ll \theta_j$.

We may consider an alternative scenario in which the initially ordered helical fields get distorted
during the energy dissipation phase, making different field directions within the bright region of 
$\theta < \Gamma^{-1}$ \cite{zhang11}.
The PA changes can naturally occur in this scenario, but when the emission duration is short,
the PA change does not necessarily occur.
Another scenario is that the GRB jets consist of multiple shells which have globally ordered
transverse (not helical or toroidal) magnetic fields with a different direction for each
shell. It has been recently claimed that such impulsive shells can be accelerated to relativistic
speeds \cite{granot12}. In this scenario also, the PA changes naturally occur for long duration bursts with 
large number of emitting shells, but do not necessarily occur for short duration bursts with
small number of emitting shells. 

\subsubsection{SR model}

The collisionless shocks formed in the jet may produce sizable magnetic fields with random directions
on plasma skin depth scales through e.g., the Weibel instability \cite{medvedev99,gruzinov99}. 
Synchrotron emission from such fields can have high $\Pi$, provided that the field directions
are not isotropically random, reflecting the direction of the shock propagation direction. 
In many studies, the extreme case is assumed, i.e.,
the field directions are confined in the plane parallel to the shock front
\cite{granot03,nakar03,toma09}. We call this the ``SR model'' (synchrotron, random-field model).

\begin{figure}[t]
\begin{minipage}{0.5\hsize}
\begin{center}
\includegraphics[scale=0.6]{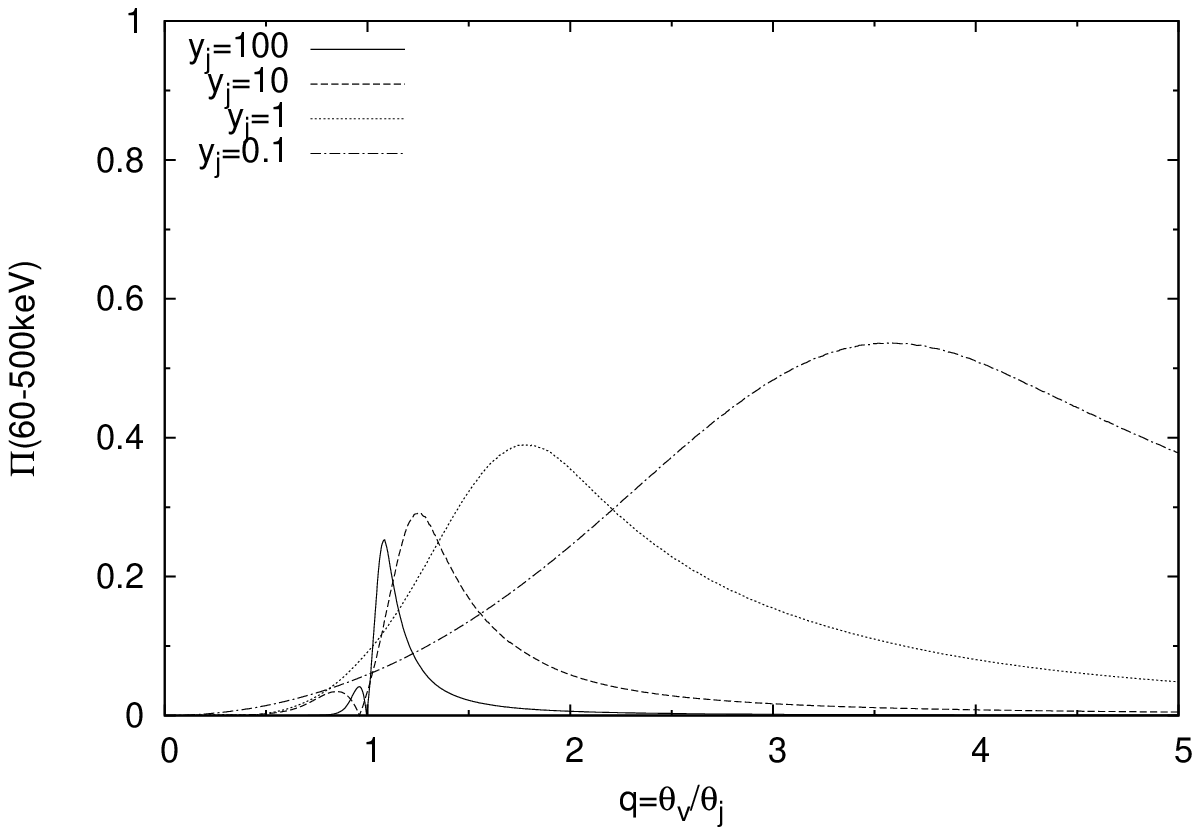}
\end{center}
\end{minipage}
\begin{minipage}{0.5\hsize}
\begin{center}
\includegraphics[scale=0.6]{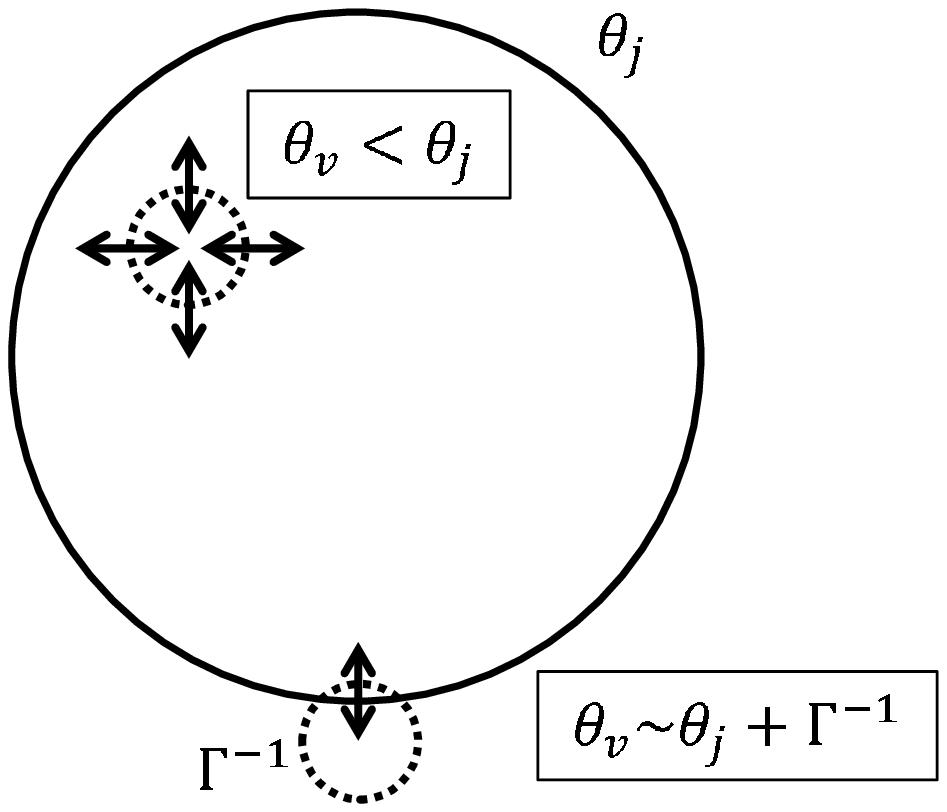}
\end{center}
\end{minipage}
\caption{
Left: Polarization degrees as functions of $q=\theta_v/\theta_j$ in the SR model
(see \cite{toma09} for details). 
Right: Schematic picture of the jet. The net polarization property is determined by the 
bright emission from the points with $\theta \sim \Gamma^{-1}$ around the line of sight,
whose polarization vectors (represented by the thick arrows) are axisymmetric.
}
\label{fig:SR}
\end{figure}

In this model, the radiation propagating in the direction parallel to the shock front is maximally
polarized in the comoving frame of the emitting fluid. Such radiation is observed as that from 
the points with $\theta = \Gamma^{-1}$ around the line of sight. As a result, the local polarization
vectors are axisymmetric around the line of sight (see Figure~\ref{fig:SR} right). 
If the jet is observed from an off-axis angle,
$\theta_v \gtrsim \theta_j$, all the polarization vectors are not canceled and the net polarization remains.
Figure~\ref{fig:SR} (left) shows $\Pi$ calculated in this model as a function of the viewing angle 
$\theta_v$ in respect of the jet axis for different values of $y_j \equiv (\Gamma \theta_j)^2$.
In this model with the uniform emissivity over the shell, a high $\Pi$ can be obtained only when
$\theta_v \sim \theta_j + \Gamma^{-1}$. For this configuration, one cannot have the PA change
even for a fixed $\theta_v$ but different $\Gamma$. For $\theta_v < \theta_j$, one can have the PA change with
varying $\Gamma$, but $\Pi$ is very low. 

The observed PA change with high $\Pi$ may suggest the patchy emission structure in this model \cite{yonetoku11}.
If the emission is patchy, $\Pi$ can be high even for $\theta_v < \theta_j$, and one can have
the PA changes (see also \cite{lazzati09}). The characteristic angular size of the patches may be
hydrodynamically constrained to be $\theta_p \gtrsim \Gamma^{-1}$. In this model, however, one requires 
fine tuning that the observed patches should be dominated by those with $\theta_{vp} \sim \theta_p + \Gamma^{-1}$ 
to have $\Pi \gtrsim 30\%$, where $\theta_{vp}$ is the viewing angle of the patch. 
The patches observed with $\theta_{vp} \lesssim \theta_p$ decrease the net $\Pi$.
On the other hand, the bursts we observed are all very bright, which implies that some patches are seen with 
$\theta_{vp} \lesssim \theta_p$. Therefore, the SR model is not favored to explain the observed $\Pi \gtrsim 30\%$.

\subsubsection{SH model}

The internal shocks may also produce strong magnetic fields with random directions on hydrodynamic
scales, much larger than the plasma skin depth scales, through e.g., the Richtmyer-Meshkov instability
\cite{inoue11,gruzinov99}. We call this the ``SH model" (synchrotron model with random fields on
hydrodynamic scales).
If the field directions are isotropically random, the net polarization degree is $\Pi \sim 
\Pi^{\rm syn}_{\rm max}/\sqrt{N}$, where $N$ is the number of independent patches with coherent
field in the bright region with $\theta \sim \Gamma^{-1}$ around the line of sight, and 
the PA change can be naturally realized. Unlike the SR model, the emission from patches seen 
with small $\theta_{vp}$ can have high $\Pi$, so that this model is in agreement with the high brightness of the bursts. 

By utilizing the MHD simulations of internal shocks with initial density fluctuations,
Inoue et al. (2011) \cite{inoue11} deduced $N \sim 10^3$ from the typical scale of the 
coherent magnetic fields, which did not appear to be
consistent with the observed $\Pi \gtrsim 30\%$. However, the recent detailed analysis of the 
numerical simulation suggests that the magnetic fields perpendicular to the shock front are 
selectively amplified, which might increase the net $\Pi$ \cite{inoue13}. The aim of this recent
simulation is to explain the radially aligned fields observed in some young supernova remnants,
e.g., \cite{reynoso13}, in which the shock velocity is non-relativistic, although probably the properties
of the amplified fields may not be different in the mildly-relativistic case like the internal shocks
of jets (T. Inoue, private communication).

\subsection{Photospheric Emission Model}

The photospheric emission model assumes that the emission at $E \gtrsim E_p$ is the quasi-thermal
radiation from the photosphere (see \cite{beloborodov11} and references therein). 
The emission at $E < E_p$ may be a superposition of many
quasi-thermal components with different temperatures \cite{ryde10,toma11,mizuta11} or contribution of
the synchrotron emission \cite{vurm11}. The quasi-thermal radiation can have high $\Pi$ when the 
radiation energy is smaller than the baryon kinetic energy at the photosphere \cite{beloborodov11}.
In this case, the angular distribution of the radiative intensity is beamed towards the expansion
direction {\it in the fluid frame} around the photosphere, and the last electron scatterings produce the 
linear polarization. Then the photons scattered to the transverse directions are maximally polarized
in the fluid frame. This situation is the same as the SR model. As a result, the observed polarization 
vectors are symmetric around the line of sight. The polarization degree of emission from a given 
direction is determined by the brightness distribution below the photosphere (the last scattering 
points are widely distributed below the photosphere) and the photon anisotropy before the last 
scatterings, which provide $\Pi \leq \Pi^{\rm qt}_{\rm max} \sim 40\%$ \cite{beloborodov11}.  

Similar to the SR model, the observed PA change may suggest the patchy emission structure. 
The angular size of the patches may be hydrodynamically constrained to be $\theta_p \gtrsim \Gamma^{-1}$.
Only the patches with $\theta_{vp} \sim \theta_p + \Gamma^{-1}$ provide a high $\Pi$ and the bright patches 
with $\theta_{vp} \lesssim \theta_p$ give $\Pi \ll \Pi^{\rm qt}_{\rm max}$. The observed high brightness of
the three bursts disfavors the quasi-thermal emission. 

However, it is possible that the synchrotron emission contributes to the net polarization at $E < E_p$
\cite{vurm11}.
One may consider the SO or SH model for this synchrotron component. This interpretation may be valid
for GRB 100826A and GRB 110721A, for which the GAP energy range is below $E_p$.

\section{Observational Test of $CPT$ Invariance}

Lorentz invariance is the fundamental symmetry of Einstein's theory of relativity. However, in quantum
gravity such as superstring theory, loop quantum gravity, and Ho\v{r}ava-Lifshitz gravity, Lorentz
invariance may be broken either spontaneously or explicitly 
(see \cite{toma12,myers03,gleiser01,mukohyama10} and references therein). 
Dark energy, if it is a rolling scalar 
field, may also break Lorentz invariance spontaneously. In the absence of Lorentz invariance, the 
$CPT$ theorem in quantum field theory does not hold, and thus $CPT$ invariance, if needed, should be
imposed as an additional assumption. Hence, tests of Lorentz invariance and those of 
$CPT$ invariance can independently deepen our understanding of the nature of space-time.

If $CPT$ invariance is broken then group velocities of photons with right-handed and left-handed circular 
polarizations should differ slightly, leading to birefringence and a gradual PA rotation of linear 
polarization. This is quite similar to the Faraday rotation effect on the photons propagating through
the ordered magnetic fields \cite{rybicki79}, although the $CPT$ invariance violation effect is stronger
in higher energy range, in the opposite way to the Faraday rotation effect.
The $CPT$ invariance violation effect is very tiny, but its accumulation from a celestial object 
through the long distances may be significant. Thus a test of such an effect can be performed with the 
polarization of the GRB prompt emission, which originates at the cosmological distances and is
bright in high-energy gamma-rays.

As we explicitly show below, the reliable measurement of gamma-ray linear polarization presented above 
enables us to obtain a strict limit on $CPT$ violation. In order to do this, source distances of the 
three GRBs are required to be estimated, but unfortunately their redshifts are not determined.
Instead, we use a well-known distance indicator for GRBs, $L_p = 10^{52.43\pm 0.33} \times
[E_p (1+z)/355\;{\rm keV}]^{1.60\pm 0.082}\;{\rm erg}\; {\rm s}^{-1}$, where $L_p$ is the peak luminosity
and $z$ is the source redshift \cite{yonetoku04,yonetoku10}. Once we measure $E_p$ and peak flux
we can calculate a possible redshift. This correlation equation includes systematic uncertainty
caused by the data scatter. Possible redshifts are then estimated to be
$0.71 < z < 6.84$, $0.21 < z < 1.09$, and $0.45 < z < 3.12$ with $2\sigma$ confidence level
for GRB 100826A, GRB 110301A, and GRB 110721A, respectively.

\subsection{Limit on $CPT$ violation}

In the effective field theory approach, Lorentz violating (LV) effects suppressed by $E/M_{\rm Pl}$
arise from dimension-5 LV operators \cite{myers03}, where $M_{\rm Pl} = (\hbar c/G)^{1/2} = 1.22 \times 10^{19}\;$GeV
is the Planck mass. 
Hereafter we shall adopt the unit with $\hbar = c = 1$. In the photon sector
they manifest as the Lorentz- and $CPT$-violating dispersion relation of the form
\begin{equation}
E_{\pm}^2 = p^2 \left( 1 \pm 2\xi \frac{p}{M_{\rm Pl}} \right),
\label{eq:dispersion}
\end{equation} 
where $\pm$ denotes different circular polarization states and $\xi$ is a dimensionless parameter.

If $\xi \neq 0$, then Eq.~(\ref{eq:dispersion}) leads to slightly different group velocities for 
different polarization states. Hence, the polarization vector of a linearly polarized wave rotates
during its propagation \cite{gleiser01}. The rotation angle in the infinitesimal time interval
$dt$ is $d\theta = (E_+ - E_-)dt/2 \simeq \xi p^2 dt/M_{\rm Pl}$. Substituting $p=(1+z)k$,
$dt = -dz/[(1+z)H]$ and $H^2 = H_0^2 [\Omega_m (1+z)^3 + \Omega_\Lambda]$, the rotation angle
during the propagation from the redshift $z$ to the present is expressed as
\begin{equation}
\Delta \theta (k,z) \simeq \xi \frac{k^2}{M_{\rm Pl} H_0} \int^{z}_0 \frac{(1+z')dz'}
{\sqrt{\Omega_m (1+z')^3 + \Omega_\Lambda}}.
\end{equation}
Here $k$ is the comoving momentum, $H_0 = 1.51 \times 10^{-42}\;$GeV, $\Omega_m = 0.27$,
and $\Omega_\Lambda = 0.73$.

For large $\Delta \theta(k,z)$ in the GAP energy range ($E_{\rm min} = 70\;$keV and
and $E_{\rm max} = 300\;$keV), the polarizations at different energies are canceled, and
the net polarization integrated over the GAP energy range is significantly depleted and 
cannot be as high as the observed level.
The detection of highly polarized gamma-ray photons in the GAP energy range thus implies that 
$|\Delta \theta(E_2,z) - \Delta \theta(E_1,z)| \leq \pi/2$, where we consider that 
a certain proportion of the total number of photons are included in $E_1 < k < E_2$. 
In order to obtain an upper bound on 
$|\xi|$ from this inequality, we set $E_1 = E_{\rm min}$ and determine $E_2$ by
$\int^{E_2}_{E_{\rm min}} E^{\gamma_{\rm B}} dE/\int^{E_{\rm max}}_{E_{\rm min}} E^{\gamma_{\rm B}} dE = \Pi$,
where $\Pi$ is the net polarization degree over the GAP energy range $E_{\rm min} \leq k \leq E_{\rm max}$.
This prescription for $E_{1,2}$ corresponds to an ideal situation in which the detected signal
has 100\% of the polarization degree and uniform polarization direction over the range 
$E_{\rm min} \leq k < E_2$, but has no polarization in the range $E_2 \leq k \leq E_{\rm max}$.
With more realistic momentum-dependencies of the polarization degree and direction, $E_2$ would 
be higher and, hence, the bound on $|\xi|$ would be tighter. 
For GRB 110721A, the $2\sigma$ lower limits $\gamma_{\rm B} > -0.98$ and $\Pi > 35\%$ for the 
GAP energy range lead to $E_2 \simeq 120\;$keV.
Setting $z>0.45$ in $|\Delta \theta(E_2,z) - \Delta \theta(E_{\rm min},z)| \leq \pi/2$, we obtain
the constraint as $|\xi| < 7 \times 10^{-15}$.

More accurate constraints are obtained by requiring that $\sqrt{Q^2+U^2}/N > \Pi$, where
$N = \int^{E_{\rm max}}_{E_{\rm min}} E^{\gamma_{\rm B}} dE$, $Q=\int^{E_{\rm max}}_{E_{\rm min}}
E^{\gamma_{\rm B}} \Pi_i \cos[2\Delta\theta(E,z)]$, and $Q=\int^{E_{\rm max}}_{E_{\rm min}}
E^{\gamma_{\rm B}} \Pi_i \sin[2\Delta\theta(E,z)]$ with the intrinsic polarization degree $\Pi_i = 1$.
Using $\Pi > 0.35$ and $\gamma_{\rm B} > -0.98$, we obtain the constraint from GRB 110721A as
\begin{equation}
|\xi| < 2 \times 10^{-15},
\end{equation}
which is tighter than the above rough estimate.
From GRB 100826A and GRB 110301A, we obtain weaker constraints \cite{toma12}.
See \cite{toma12} for comparison of this result to the constraints from other experiments.

\section{Summary and Discussion}

As for the prompt emission mechanism,
the SO model can explain all of the GAP results, although even this model does not easily produce
polarization as high as $\Pi > 40\%$.
The SR model requires the off-axis viewing angles of the emission regions, for which the brightness
is low because of the relativistic beaming effect. The observed high brightness of the three bursts
disfavors this model. 
The SH model is also disfavored if the turbulent field directions are isotropic and $N \sim 10^3$
as suggested by the MHD simulations of \cite{inoue11}. However, the field directions might be 
anisotropic as claimed by the recent detailed analysis of the simulations \cite{inoue13}. 
More careful studies are needed.
The polarization properties of the quasi-thermal emission from the photosphere 
are quite similar to those of the SR model, and thus it is also disfavored. However, it is possible 
that the contribution of the synchrotron emission from the photosphere at $E<E_p$ \cite{vurm11} based on the 
SO or SH model might reproduce the observational results.

In general, the ordered helical magnetic fields are advected from the
central engine. If the SR, SH, or quasi-thermal model is valid, the energy density of such an ordered field component is 
sub-dominant, while if the SO model is valid, it is dominant and also controls the dynamics. 
In other words, the SO model implies that the jets are dominated by the Poynting flux rather than the 
baryon kinetic energy flux, and they are driven electromagnetically rather than by
the thermal pressure.

However, it should be noted that the SO model assumes globally ordered fields in the emission region. 
On the other hand, the observed dim X-ray afterglows just after the prompt emission (i.e., the 
so-called X-ray plateau) have suggested that prompt emission has very high efficiency
(even $>90\%$ for some bursts) \cite{ioka06,zhang07}, which means that the energy dissipation,
usually involving field distortion, occurs globally. Reconciling the high $\Pi$ with the high 
radiation efficiency looks to be a dilemma, which will have to be resolved in more quantitative modeling.

Anyway, more accurate observational data and the statistical study of such data are needed and essential 
to solve the problems, cf., \cite{toma09}. There are several polarimeter mission concepts, such as 
PETS \cite{mcconnell13} and TSUBAME \cite{yatsu12}.
There is a possibility that some bursts would be detected more reliably with $\Pi \gg 40\%$, which 
suggests the Compton drag model \cite{eichler03,lazzati04,toma09}. 
The polarization spectrum would be also valuable. The composite model of the quasi-thermal
plus synchrotron emissions \cite{vurm11} may show a specific tendency in the polarization spectra.

We have obtained the tight constraint on the $CPT$ invariance parameter, $|\xi| < 2 \times 10^{-15}$,
from the reliable measurement of the gamma-ray polarization with GAP \cite{toma12}. Tighter constraint, 
$|\xi| < 3 \times 10^{-16}$, has been recently obtained for GRB 061122 with spectroscopic redshift
by using the INTEGRAL-IBIS data \cite{gotz13}. 

In the effective field theory approach \cite{myers03}, there is only one operator that leads to a 
linear energy dependence of the speed of light in vacuum, and it is the dimension-5 $CPT$-odd LV
operator considered above (see Eq.~\ref{eq:dispersion}). 
Constraints on the same operator from observation of energy dependence
of GRB light curve, as performed for GRB 090510 \cite{abdo09}, are much weaker than those
from observation of polarization such as ours. For this reason, it is natural
to interpret the light-curve observation as limits on the dimension-6 LV operator. 
The observation of GRB 090510 puts the lower bound on the quantum gravity mass scale as 
$M_{QG,2} > 10^{11}\;$GeV. This is consistent with the natural expectation that the quantum gravity
mass scale is of the order of the Planck mass.

\Acknowledgements
I am grateful to D. Yonetoku, T. Murakami, S. Gunji, T. Mihara, S. Mukohyama, T. Inoue for useful discussions. This work is partly supported by JSPS Research Fellowships for Young Scientists
No. 231446.

\end{document}